\documentclass[notitlepage,aps,pra,twocolumn,groupaddress,10pt]{revtex4-2}
\usepackage[english]{babel}
\usepackage{array}% authblk
\usepackage{stmaryrd}% authblk
\usepackage{amssymb,amsmath,amsthm,amsfonts,amsbsy}
\usepackage{bm,bibunits,color,chngcntr,epsfig,epstopdf,graphicx,dsfont}
\usepackage{lipsum,makecell,mathrsfs,rotating}
\usepackage{hyperref}% authblk

\usepackage[normalem]{ulem}
\usepackage[absolute]{textpos}

\newcommand{\be}{\begin{equation}}
\newcommand{\ee}{\end{equation}}
\newcommand{\bea}{\begin{eqnarray}}
\newcommand{\eea}{\end{eqnarray}}
\newcommand{\ket}{\rangle}
\newcommand{\bra}{\langle}

\newcommand{\I}{\mathds{1}}
\newcommand{\ra}{\rightarrow}

\def\C#1{\mathcal #1}

\def\Tr#1{\text{tr} #1}

\definecolor{gray}{gray}{0.9}

\begin{document}
\newtheorem{theorem}{Theorem}
\newtheorem{prop}[theorem]{Proposition}
\newtheorem{corollary}[theorem]{Corollary}
\newtheorem{open problem}[theorem]{Open Problem}
\newtheorem{conjecture}[theorem]{Conjecture}
\newtheorem{definition}{Definition}
\newtheorem{remark}{Remark}
\newtheorem{example}{Example}
\newtheorem{task}{Task}

\title{Quantum simulation in the entanglement picture}
\author{Dong-Sheng Wang}
\email{wds@itp.ac.cn}
\affiliation{Institute of Theoretical Physics, Chinese Academy of Sciences, Beijing 100190, China \\
School of Physical Sciences, University of Chinese Academy of Sciences, Beijing 100049, China}
\author{Xiang Xu}
\affiliation{Institute of Theoretical Physics, Chinese Academy of Sciences, Beijing 100190, China \\
School of Physical Sciences, University of Chinese Academy of Sciences, Beijing 100049, China}
\author{Yuan-Dong Liu}
\affiliation{Institute of Theoretical Physics, Chinese Academy of Sciences, Beijing 100190, China \\
School of Physical Sciences, University of Chinese Academy of Sciences, Beijing 100049, China}

\begin{abstract}
The notion of ``picture'' is fundamental in quantum mechanics.
In this work, a new picture, which we call entanglement picture,
is proposed based on the novel channel-state duality, whose importance 
is revealed in quantum information science. 
We illustrate the application of entanglement picture in quantum algorithms 
for the simulation of many-body dynamics, quantum field theory, thermal physics,
and more generic quantities. 
\end{abstract}

\maketitle

\section{Introduction}

Quantum entanglement is a fundamental concept in modern quantum physics~\cite{HHH09}.
In recent decades, it has generated profound development, 
ranging from the field of quantum information~\cite{NC00},
the spacetime geometry~\cite{RT06}, 
to the study of topological order and quantum phase transition~\cite{ZCZ+15}.

Quantum entanglement leads to an important way of describing 
quantum state, which is matrix-product state (MPS)~\cite{PVW+07}.
Any pure finite-dimensional quantum state, regardless of spatial geometry, 
can be written as a MPS
by expressing its amplitudes as 
$\psi_{\vec{i}}=\text{tr}(B A_{\vec{i}})$
for matrices $A$ and $B$ acting on a so-called bond space.
The operator $B$ specifies the boundary condition,
and a collection of $As$ for each site forms a quantum channel.
A central feature it captures is the bulk-edge duality, namely,
the bulk feature can be captured by 
the features of the channels acting on the bond space,
which is the space for the edge mode.

The study of the bulk-edge duality is the focus of this work.
We first find that, using the language from quantum information theory,
the bulk-edge duality can be traced back to 
a fundamental principle,
which is the channel-state duality~\cite{Cho75,Jam72}. 
It is a type of space-time duality that converts
a dynamics into a state, from which the original dynamical feature
can be recovered.

%We also find that it can also explain the Suzuki quantum-classical duality
%together with Wick's rotation mapping between temperature and time.

Based on the channel-state duality, 
we introduce the \emph{entanglement picture}, 
which studies a quantum system from its entanglement. 
Recall that in quantum mechanics, 
there are so-called ``pictures'' 
and usually there are three of them: 
Schr\"odinger picture, Heisenberg picture, and the interaction picture. 
A picture is a perspective from which to describe a quantum system. 
These pictures are equivalent with respect to observable effects
\be \Tr(O\rho_t)=\Tr(OU\rho U^\dagger)= \Tr(U^\dagger O U\rho )=\Tr(O_t \rho) \ee 
for any observable $O$, unitary evolution operator $U$, and state $\rho$.
For $U=e^{-itH}$ with a Hamiltonian $H=H_0+V$, 
the interaction picture considers evolution relative to $U_0=e^{-it H_0}$,
the effective Hamiltonian is $V_I=U^\dagger_0VU_0$ which generates $U_I$.
With $\rho_I=U^\dagger_0\rho U_0$, then
\be \Tr(OU\rho U^\dagger)= \Tr(OU_I\rho_I U_I^\dagger). \ee 
The evaluation of expectation values boils down to that of overlap 
$\bra \psi|\phi\ket$ between states.
Feynman's path integral may also be viewed as a picture as 
it can compute overlap (such as propagator) in a novel way
based on Lagrangian and action~\cite{Fey48}.

The Schr\"odinger picture describes the evolution of a state, 
but it does not specify what changes, 
the basis states $|\vec{i}\ket$ or the amplitudes $\psi_{\vec{i}}$, however. 
Usually, it refers to the change of basis by acting $U$ on the basis states $|\vec{i}\ket$,
while it can also be equivalently treated as the change of amplitudes.
The entanglement picture instead focuses on the change of amplitudes,
and it in particular relies on the formalism of MPS. 
An overlap $\bra \psi|\phi\ket$ for pure states $\psi,\phi$ living in the physical space
is computed from other overlaps $\bra \mu|\nu\ket$
computed in the entanglement picture
for pure states $\mu,\nu$ living in the entanglement space.
Therefore, the expectation value has a new form
\be \Tr(O\rho_t)= \Tr(\hat{O}\hat{\rho}_t) \ee 
for $\hat{O}$ and $\hat{\rho}_t$ as observable and state in the entanglement picture.
A brief comparison of these pictures is presented in Table~\ref{tab:pic}.

% \begin{table}[b!]
%     \footnotesize  \centering
%     \begin{tabular}{|p{1.8cm}|p{1.6cm}|p{1.8cm}|p{1.6cm}|p{1.9cm}|}
% %\begin{tabular}{|c|c|c|c|c|}   
% \hline
% \textbf{Picture} & \textbf{State} & \textbf{Operator} & \textbf{Evolution} & \textbf{Key Feature} \\
% \hline
% Schr\"odinger & evolves & static & unitary & state-centric \\ \hline
% Heisenberg & static & evolves & Heisenberg eq. & operator-centric \\ \hline
% Interaction & evolves (interaction) & evolves (free) & interaction Hamiltonian & perturbation theory \\ \hline
% Entanglement (this work) & mapped to channel network & mapped to measurements & channel network & evolution of entanglement structure \\
% \hline
% \end{tabular}   
% \caption{A comparison of the pictures in quantum mechanics including 
%     the new entanglement picture proposed in this work. }
%     \label{tab:pic}
% \end{table}

\begin{table}[b!]
    \footnotesize  \centering
    \begin{tabular}{|c|c|c|c|c|}
    \hline
    \textbf{Picture} & \textbf{State} & \textbf{Operator} & \textbf{Evolution} & \textbf{Key Feature} \\
    \hline
    \parbox{1.8cm}{\centering Schr\"odinger} & 
    \parbox{1.6cm}{\centering evolves} & 
    \parbox{1.8cm}{\centering static} & 
    \parbox{1.6cm}{\centering unitary} & 
    \parbox{1.9cm}{\centering state-centric} \\ \hline
    \parbox{1.8cm}{\centering Heisenberg} & 
    \parbox{1.6cm}{\centering static} & 
    \parbox{1.8cm}{\centering evolves} & 
    \parbox{1.6cm}{\centering Heisenberg eq.} & 
    \parbox{1.9cm}{\centering operator-centric} \\ \hline
    \parbox{1.8cm}{\centering Interaction} & 
    \parbox{1.6cm}{\centering evolves (interaction)} & 
    \parbox{1.8cm}{\centering evolves (free)} & 
    \parbox{1.6cm}{\centering interaction Hamiltonian} & 
    \parbox{1.9cm}{\centering perturbation theory} \\ \hline
    \parbox{1.8cm}{\centering Entanglement (this work)} & 
    \parbox{1.6cm}{\centering mapped to channel network} & 
    \parbox{1.8cm}{\centering mapped to measurements} & 
    \parbox{1.6cm}{\centering channel network} & 
    \parbox{1.9cm}{\centering evolution of entanglement structure} \\ \hline
    \end{tabular}   
    \caption{A comparison of the pictures in quantum mechanics including 
    the new entanglement picture proposed in this work. }
    \label{tab:pic}
\end{table}

The MPS formalism and its tensor-network extensions 
are often used as classical algorithms to study quantum systems~\cite{PKS19}. 
Here we develop the entanglement picture which is quantum
and can be used to design quantum algorithms. 
The problem we need to solve is that 
there is a mismatch between the quantum circuit form which acts
on the physical space and 
the MPS form whose tensors acts on the entanglement space.
We solve this problem by employing the channel-state duality
which maps the space direction to time,
and maps the circuit evolution direction to space.

The idea of entanglement picture can be viewed as 
an extension of a few ideas in literature.
It can be dated back to the theory of valence-bond solid~\cite{AKLT87},
which uses ``bonds'' to describe the entanglement between physical particles.
In measurement-based quantum computing~\cite{RB01},
resource states can be expressed as MPS form,
and computation by on-site measurement can be described in a ``virtual picture''.
Gapped one-dimensional quantum many-body systems 
are well described by MPS of small entanglement~\cite{ZCZ+15}.
For topological order, 
non-abelian anyon braiding can be described as 
matrix-product unitary operators~\cite{PKS19}.
Different from existing schemes,
the central part of entanglement picture is the way it describes dynamics, i.e.,
it uses quantum channels to describe evolution of MPS.
We show that using quantum channels acting on the ``entanglement system'' 
can recover observable of the original system, 
hence justifying the entanglement picture.

% Some of relevant ideas are:
% \begin{itemize}
%     \item MBQC: as explained above, it only uses on-site local measurement,
%     so our entanglement picture is an extension of it;
%     \item Matrix-product unitary (MPU) operator: 
%     MPU can describe anyon braiding, it is nonlocal; 
%     the entanglement picture uses semi-local gates. 
%     \item Feynman path integral can also be viewed as a picture, 
% widely used in quantum field theory.
% It also motivates ``consistent history'' approach that applies 
% to systems with discrete variables. 
% \end{itemize}

%a ``physical picture-virtual picture'' method for MBQC, but not for evolution,
%and this does not take ``picture'' as a general methodology. 

We illustrate the usage of entanglement picture 
by showing how to compute $\bra \phi|A|\psi\ket$ 
for an arbitrary pair of states and an operator.
This covers a few physical settings in quantum simulation,
including quantum many-body dynamics~\cite{BACS07}, 
quantum field theory~\cite{JLP12},
and thermodynamics~\cite{BBM20}. 
Quantum simulation is motivated by the advantage to 
simulate quantum many-body system compared to classical simulation. 
For a task with initial state $|\psi\ket$, evolution $U=e^{-itH}$,
and measurement of $A$,
in the entanglement picture 
it uses MPS form of states and a channel network for the evolution,
and it computes the final value $\bra \psi|U^\dagger A U|\psi\ket$.
In all, the entanglement picture (EP) is not merely a reformulation 
but shifts the perspective from physical space evolution to entanglement space evolution,
enabling a modular, parallelizable approach that naturally accommodates measurement-based and
distributed quantum computing.

%[here need to survey other works on the three topics.]
%methods to compute $\bra \psi| H^n |\psi \ket$ includes~\cite{WLL20,CPB21,Sha22,LLZ23,ZZS25},
%relate to thermal quantity, ground state. 

% The entanglement picture (EP) is universal, i.e., 
% it can be applied to simple system, or system with 
% exponential growth of Hilbert space dimension, i.e. many-body system. 

% So quantum simulation is a proper setting to use EP, 
% and it yields a novel type of quantum simulation algorithm. 

% These quantum simulation algorithms have lower cost 
% compared with direct algorithms that do not take advantage of the duality.  

% The formalism can also be extended.
% We show that, given access to the entanglement system,
% we can use quantum superchannels to generate more general dynamics on a quantum system.
% This also includes the usual case as special cases.
% Other techniques can be used to optimize some features of the quantum simulation, 
% such as using OQT and OQC. 
% The quantum channel simulation has been studied. 
% This is an application of QvN model of quantum computing.

%To find ground state, we need to use variation techniques, can use Wick rotation.

\section{Channel-state duality}

The channel-state duality and MPS are historically developed 
in different contexts, 
and here we show that the bulk-edge duality for MPS
originates from the channel-state duality.

% The channel-state duality relies on the map 
% \be |i\ket \bra j| \mapsto |i,j\ket \ee 
% It is a space-time map. It maps time to space. 
% It is linear and reversible but not unitary since it changes the Hilbert space dimension.
% The inverse is 
% \be |i,j\ket  \mapsto |i\ket \bra j| \ee 

Quantum dynamics is in general described as completely positive, trace-preserving mappings,
or known as quantum channels~\cite{NC00}. 
This actually includes state preparation, unitary evolution, and measurement
as special cases. 
Given a channel $\Phi: \C D(\C H_1) \ra \C D(\C H_2)$ from an input system 
$\C H_1$ to an output system $\C H_2$,
it can be represented as a state 
\be \omega_{\Phi} := \Phi \otimes \I (\omega)=\frac{1}{d}\sum_{ij} 
\Phi(|i\ket \bra j|)\otimes |i\ket \bra j|, \ee
known as a Choi state~\cite{Cho75,Jam72},
for $\omega:=|\omega\ket\bra \omega|$ as a Bell state, $d=\text{dim} \C H_1$.
From Stinespring's dilation, 
the channel $\Phi$ can also be represented as a unitary $U$
requiring an ancilla initialized at $|0\ket$ such that
$\Phi(\rho)=\text{tr}_a\left(U(\rho\otimes |0\ket\bra 0|) U^\dagger \right)$
for the trace over ancilla~\cite{NC00}. 
With the ancilla instead,
the resulting tripartite state 
is a purified Choi state, $|\phi_\Phi\ket$,
which is a purification of the Choi state $\omega_{\Phi}$.

% Recall that the channel-state duality~\cite{Cho75,Jam72} states that a channel 
% $\Phi: \C D(\C H_1) \ra \C D(\C H_2)$ 
% can be represented as a Choi state 
% \be \omega_{\Phi} := \Phi \otimes \I (\omega)=\frac{1}{d}\sum_{ij} 
% \Phi(|i\ket \bra j|)\otimes |i\ket \bra j|, \ee 
% for $\omega:=|\omega\ket\bra \omega|$ as the Bell state.
% It is clear to see the partial trace over the input and output space  
% yields $\text{tr}_1 \omega_{\Phi}=\Phi(\I) /d$,
% and $\text{tr}_2 \omega_{\Phi}= \I /d$,
% for $d$ as the input space dimension. 
% Therefore, Choi states form a convex subset 
% $\C C(\C H_1\otimes \C H_2) \subset \C D(\C H_1\otimes \C H_2)$
% of the set of bipartite states. 
% The Kraus operators can be found from the eigenvalue decomposition of $\omega_{\Phi}$,
% and the rank of the channel is the rank of $\omega_{\Phi}$.

% After a superchannel, we have 
% \be \hat{\C S} (\Phi)(\rho)=\text{tr}_{\bar{\text{A}}} \C V \otimes \tilde{\C U} 
% (\omega_{\Phi} \otimes \omega) (\I\otimes \rho^t\otimes |0\ket \bra 0|), \ee
% where the support of each operator shall be easy to see hence omitted for simplicity.
% The trace is over the subsystems except the top one, A.
% The unitary $\tilde{\C U}$ is the transpose of $U$ conjugated by a swap. 

Given a Choi state $\omega_{\Phi}$, 
the inverse map is to recover its action on state $\Phi(\rho)$.
This can be achieved as 
\be \Phi(\rho)= d \; \text{tr}_1 [\omega_{\Phi} (\I \otimes \rho^t) ], \label{eq:readout}\ee
for $\rho^t$ as the transpose of a state $\rho \in \C D(\C H_1)$
and the trace is on the input space $\C H_1$.
The duality maps state preparation to measurement,
and $\rho^t$ can be realized as a binary measurement 
\be \{M_0=\sqrt{\rho^t}, M_1=\sqrt{\I-\rho^t}\}, \label{eq:isi} \ee 
which guarantees the correct expectation value 
$\text{tr}(\C A \Phi(\rho))$ for any observable $\C A$ 
from the output space $\C H_2$~\cite{WC20,WQ22}.
Note for the outcome 1, 
the offset $\text{tr}(\C A \Phi(\I))/d$ can be removed to get the correct value.
This completes the description of the channel-state duality. 

% and the final output in terms of an 
% expectation value $\text{tr}(\C A \Phi(\rho))$ for observable $\C A$ 
% can be read out from the output space~\cite{W20_choi,W22_qvn}.
%Note the superscript $t$ is the transpose. 
%This leads to the universality to simulate any quantum algorithm 
%in the usual circuit model whose input needs to be prepared first~\cite{W22_qvn}. 

% For the outcome 0, the correct answer is obtained. 
% For the outcome 1, the offset $\text{tr}(\C A \Phi(\I))/d$ can be deleted to get the correct result.
% This is referred to as the initial-state injection (ISI) scheme. 

\section{Matrix-product state}

For a $N$-body quantum system with local dimension $d_n=\text{dim} \C H_n$, $n\in [1,N]$,
a pure state $|\psi\ket\in \bigotimes_n \C H_n$ 
can be written as a matrix-product state (MPS) 
\be |\psi\ket= \sum_{i_1 \cdots i_N} \Tr(B A_{i_N}\cdots A_{i_1}) |i_1 \cdots i_N\ket, \ee 
for an edge operator $B$ and bulk operators $\{A_{i_n}\}$
acting on the so-called ``bond'' space, or entanglement space~\cite{PVW+07}.
The MPS formalism is a proper way to describe entangled state, 
and the ``entanglement'' we referred to in this work is the Schmidt rank,
which is a measure of bipartite entanglement via the bond space dimension~\cite{HHH09}.
The state $|\psi\ket$ can be normalized by noting $\bra \psi|\psi\ket=\Tr \left( \C M_1 \cdots \C M_N (B\otimes B^*) \right)$ for each transfer operator
\be \C M_n:=\sum_{i_n} A_{i_n} \otimes A_{i_n}^*, \ee 
which actually is a representation of the channel $\Phi_n$ formed by the set of Kraus operators $\{A_{i_n}\}$
for each site $n$. 
In this form, an evolution $\Phi(\rho)$ is expressed as $\C M|\rho\ket$
for the reshaping $|\rho\ket=\sum_{ij}\rho_{ij} |i,j\ket$ and $\rho=\sum_{ij}\rho_{ij} |i\ket \bra j|$~\cite{BZ06}.
For an observable $O_1\otimes \cdots \otimes O_N$,
its expectation value is 
\be \bra \psi|O_1\otimes \cdots \otimes O_N |\psi\ket =
\Tr(\hat{O}_1\cdots \hat{O}_N (B\otimes B^*)) \label{eq:mpsduality}\ee 
for $\hat{O}= \C M_O=\sum_{ij} \bra i|O|j\ket A_i \otimes A_j^*$
as the representation of an operator $O$ relative to a channel at each site.
This is often referred to as the bulk-edge duality (or correspondence)
as the static bulk property of $|\psi\ket$ is equivalent to 
the dynamic property of the edge system.

To illustrate this, 
it is not hard to see an operator $\hat{O}$ reduces to a measurement,
and an operator $\C M$ is `free' evolution of a channel $\Phi$.
Each local observable $O$ decomposes into a set of eigenstates,
and we only need to consider the transfer operator $\C M_\eta= D\otimes D^*$
for each of them $|\eta\ket$ with 
$|\eta\ket=\sum_i \eta_i |i\ket$ and $D=\sum_i \eta_i A_i$.
An operator $D$ can be further reduced to its eigenstates,
and eventually the value (\ref{eq:mpsduality}) reduces to the computation 
of overlaps. % of the form $\bra d|\rho|d\ket$. 
For instance, for a translation-invariant system with
an open-boundary condition $B=|\ell\ket \bra r|$,
a two-body observable $\bra O_x O_y\ket$ at sites $x$ and $y>x$
reduces to the sum of values 
\be \bra \eta_x |\Phi^{(x-1)}(\ell)|\eta_x\ket  
    \bra \eta_y |\Phi^{(y-x)}(\eta_x)|\eta_y\ket 
    \bra r |\Phi^{(N-y)}(\eta_y)|r\ket \ee 
for $\eta_x$ and $\eta_y$ as local projectors whose form 
follows from our analysis above. 
It is clear that each overlap of the form $\bra d|\rho|d\ket$
is simulated by a quantum channel dynamics followed by a measurement. 

% Note $O$ does not need to be Hermitian, 
% and the only requirement is that it can be diagonalized.

As our first result,
it is clear to see the bulk-edge duality originates from the channel-state duality
from our description above. 
A state in the MPS form, ignoring the edge operator, 
is a purified Choi state $|\phi_{\Phi_N \cdots \Phi_1}\ket$.
Observable, as well as the edge operator, introduces  
initial state and measurement for each segment of evolution. 
This fact is simple but crucial,
as we will show below the bulk-edge duality also extends to dynamics,
leading to the entanglement picture.

\section{Entanglement picture}

A few methods have been developed to study dynamics of MPS~\cite{PKS19}.
A unitary evolution $U$ can be expressed as a brickwork quantum circuit 
\be U=\prod_l U_l, \; U_l=\otimes_n u_{n,n+1} \label{eq:brick} \ee 
for each layer $U_l$ 
as a transversal product of nearest-neighbor two-local gates $u_{n,n+1}$.
%Such circuits are also known as cellular automata~\cite{} 
%and are universal for quantum computing. 
Each local gate $u$ acts on the physical sites,
and the task is to convert its action to the entanglement space.
One method is to perform singular-value decomposition (SVD)
for the local tensor $A_{i_{n+1}}A_{i_n} u_{n,n+1} |i_n,i_{n+1}\ket$,
and this will become $B_{i_{n+1}}B_{i_n} |i_n,i_{n+1}\ket$.
Usually truncation is used to reduce the bond dimension 
by ignoring small singular values while sacrificing simulation accuracy~\cite{Vid04}. 
Another method is to use the matrix-product unitary form
of layers of gates~\cite{PKS19}, and then perform tensor contraction.

% Traditional way to deal with MPS dynamics is to apply a quantum circuit on a MPS. 
% A quantum circuit can be formed as layers of two-body local gates.
% Such circuits are also known as ``brickwork'' circuits.
% In the first layer, a local gate $U$ acts on two tensors, namely 
% \be  A_j A_i U|ij\ket \ee 
% A way to calculate this is first write this as a whole tensor $B_{ij}$,
% and then do SVD, and this will convert it into  
% \be  B_j B_i |ij\ket \ee 
% The bond dimension will usually increase as more gates are being applied.
% Truncation is used to reduce the bond dimension by 
% ignoring some singular values of small values.

% Another way is to express two layers of gates as a MPU, 
% matrix-product unitary, 
% and the basic step is also SVD. 
% The rest is tensor contraction (TC).

% Both SVD and TC are suitable for classical computers, 
% so this forms an important type of classical simulation algorithm 
% to study quantum systems. 

% The TC method is classical simulation algorithm, 
% it is not entanglement picture since it is not quantum. 

\begin{figure}[t!]
    \centering
    \includegraphics[width=0.45\textwidth]{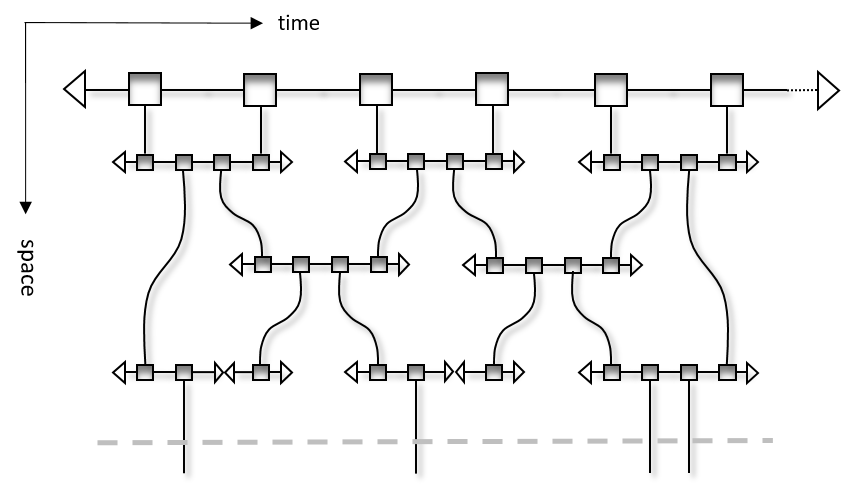}
    \caption{Schematic to illustrate the entanglement picture. 
    It shows a state $|\psi\ket$ of six explicit sites and open boundary condition,
    and a circuit $U$ with three layers of two-local gates, 
    each of which is expressed as a MPS with open boundary condition.
    The bottom dashed line indicates the conjugate of $U|\psi\ket$.
    An observable, say, $O_2 O_4$, induces the projectors inserted in between
    the channel evolution at the bottom layer.
    A vertical wire that connects two channels implies a contraction, 
    which is a projection 
    $|\omega\ket \bra \omega |$ on the two ancillary spaces. 
    A vertical wire that crosses the dashed line implies 
    the trace over it.
    }
    \label{fig:ep}
\end{figure}

Instead of using tensor contraction, 
we employ quantum channels to describe the evolution of MPS.
This involves a space-time map and naturally switches to 
the entanglement picture.
Each local gate $u$ would introduce qubits to the entanglement space.
Instead of treating $u$ as tensors,
we convert it to quantum channels by firstly mapping 
a two-local $u$ to a Choi state 
$|u\ket=\sum_{ijkl}u_{ijkl}|ijkl\ket$,
which can be written as a MPS based on SVD, 
and then mapping it back to an operator. 
See Fig.~\ref{fig:ep} and we assume open boundary condition for simplicity. 
The primary goal is to compute 
$\bra \psi| U^\dagger (O_1 \otimes \cdots \otimes O_N) U |\psi\ket$,
which is computed by a channel network interleaved with projectors 
for initial states and measurements.

A nontrivial part in a channel network is the vertical ``wires'' 
that each connect two channels. 
A vertical wire is a projection $|\omega \ket \bra \omega |$
on the two ancillary indices for two channels.
This does not require post-selection actually, 
and from (\ref{eq:isi}), the duality ensures that 
the binary measurement 
$\{|\omega \ket \bra \omega |, \I- |\omega \ket \bra \omega |\}$
will generate the correct result in a heralded way.
To see this, denote $\rho$ as a state before the measurements, 
and each measurement outcome $\otimes_i P_i$ will generate a
probability $\Tr(\otimes_i P_i \rho)$,
all of which are equivalent, for each $P_i$ as $|\omega \ket \bra \omega |$
or its complement $\I- |\omega \ket \bra \omega |$. 
That is, although the total number of measurement outcomes 
increases exponentially with the number of gates in the circuit,
all outcomes work. 
Note that more samples are needed in order to correct the offset of probability
for each case. 
This also works to deal with the boundary conditions,
e.g., for $B=|\ell\ket \bra r|$ in the initial MPS, 
a measurement $\{|r \ket \bra r|, \I- |r \ket \bra r|\}$ suffices. 
This completes the basic content of the entanglement picture.

%A channel may not preserve dimension, this is not a problem. 

%There are projections that need post-selection. 

In all, the entanglement picture shift the time evolution of states in Schr\"odinger picture 
to its entanglement. 
The channel network acts on the entanglement space. 
Given a system of size $N$ and (maximal) local dimension $d$, 
and a brickwork circuit of layers $L$, 
the entanglement picture would require about $\lfloor \frac{N}{2} \rfloor$ qudits for 
an initial MPS, and $6M$ qudits for the evolution,
where the factor 6 accounts for the number of qudits for each local gate $u$,
and $M:=L\lfloor \frac{N}{2} \rfloor$ is the total number of gates.
Modular constants, the space-time cost in the entanglement picture
is equivalent to that in Schr\"odinger picture, which is $\C O(MN)$.  
In addition, it is possible to go beyond the unitary case:
one can use generic channels to generate local evolution acting on an initial mixed state,
thus enabling open-system dynamics that can be simulated on quantum computers.
Due to purification and dilation~\cite{NC00},
this can be realized by the unitary setting in principle.

Furthermore, we observe that the circuit depth for each channel sequence for each $u$
is small, while the circuit depth for the initial edge system is large, $\sim N$.
This could be a problem if decoherence error exists,
and this can be resolved by a recent technique called oblivious quantum teleportation (OQT)~\cite{XLS25}.
The underlying idea still originates from the channel-state duality, 
namely, first break the initial MPS into segments of two tensors
and prepare the corresponding purified Choi states, $|\phi_{\Phi_{n+1}\Phi_n} \ket$,
and then perform the binary Bell measurement 
$\{|\omega \ket \bra \omega |, \I- |\omega \ket \bra \omega |\}$ 
to connect them. 
This is exactly the same measurement for the vertical connections among channels, 
but here for the horizontal connections, it works differently. 
For the outcome $|\omega \ket \bra \omega |$, it acts as identity channel,
while for the outcome $\I- |\omega \ket \bra \omega |$,
it acts as a channel 
\be \C P(\rho)=\frac{d^2}{d^2-1}\Delta (\rho)-\frac{1}{d^2-1}\rho \label{eq:dep}
\ee
for $\Delta (\rho)=\I/d$ as the completely depolarizing channel. 
The offset due to $\Delta (\rho)$ can be easily dealt with
for computing observable. 
With this technique, now the entanglement picture can be roughly described 
as an array of short channel sequences, 
with vertical connections required by the evolution,
and horizontal connections at the top layer required by the initial state. 
Besides, if it starts from an initial product state $|\psi_1\ket\cdots|\psi_k\ket$,
the high-depth of an initial MPS $|\psi\ket$ can be treated as the yield 
from a local circuit applied on a product state.
This would avoid the signal decay from Eq.~(\ref{eq:dep}) due to the usage of OQT~\cite{XLS25}.

\section{Quantum simulation algorithm}

Now we consider quantum simulation tasks as 
the application of the entanglement picture (EP), 
which can reveal more features of it.  
Such a type of simulation has been anticipated as a weak quantum simulation~\cite{W15}, 
which focuses on the computation of observable, 
instead of reproducing state preparation and evolution.

\subsection{Quantum many-body system dynamics}

Consider a local quantum Hamiltonian 
\be H=\sum_r H_r \label{eq:ham} \ee for each $H_r$ 
acting on a constant number of local sites, 
and there is a polynomial number of local terms.
The Hamiltonian quantum simulation task is to realize $U(t)=e^{-itH}$
on an initial state $|\psi\ket$.
Many methods have been developed, and here we survey two of them. 
A primary method is to apply Trotter-Suzuki decomposition to realize $U(t)$
as a sequence of local terms $e^{-i \tau H_r}$ for various short period of time $\tau$~\cite{BACS07}. 
The circuit depth of this method, however, scales as $\C O(\frac{1}{\epsilon})$ 
for $\epsilon$ as the Trotter-Suzuki accuracy. 
An exponential improvement of accuracy can be achieved with other methods, 
e.g., the linear combination of unitary algorithm~\cite{BCC+14},
which, on the other hand, would require a large ancillary control system
and also the amplitude amplification algorithm~\cite{BHM02}. 

The EP method uses a channel network. 
As it focuses on observable instead of the states, 
there is no need to maintain the coherence of the whole network all the times,
and a parallelism can be used. 
Namely, each local patches of channels can be run in parallel,
and it is a product state $\otimes_i \rho_i$ 
before making the vertical and horizontal connections, 
and the projections for the boundary conditions.
The projections are also of product form, $\otimes_j P_j$. 
Denote the underlying lattice as $\C L$, 
for each site of it corresponding to a channel, initial state or final measurement. 
A small region $\Lambda \subset \C L$ in the network
can be chosen to compute the probability value 
\be p_\Lambda=\otimes_{ij} \Tr(\rho_i P_j), \ee 
for $i,j\in \Lambda$.
The network can be divided into non-overlapping regions $\bigcup_n \Lambda_n$,
for each of which a probability $p_n$ can be computed. 
Then the projections across the regions are made to compute the final result. 
In particular, the merit to compute local values $p_n$ first,
compared with a direct scheme which is to apply all projections in parallel, 
is that they
can be used as baseline to increase the simulation accuracy.

The main cost is the cost for simulating the dynamics $U=e^{-itH}$. 
As we use Trotter-Suzuki decomposition, 
the spacetime cost is in the same order as the usual simulation algorithm 
run in the circuit model, which is also in the Schr\"odinger picture. 
It has an additional sampling cost due to the usage of measurements,
which is $\C O(N^2ML)$.
Further sampling is needed to estimate expectation value
after simulating a certain dynamics,
and such cost follows from the standard Chernoff bound~\cite{Hay17}.
Compared with other methods we mentioned above,
the EP method has a weaker requirement for maintaining coherence.
Also note that the ability to compute local values of overlaps 
does not yield an efficient classical algorithm, 
since eventually the whole channel network must be run 
to generate the whole global value of 
measurement outcomes of observable. 
It is well established that, as a seminal Lieb-Robertson bound, 
entanglement increases linearly with time for generic non-equilibrium dynamics~\cite{GKS05},
rendering classical simulation hard.

Our description of EP in the last section does not require a geometry in space, 
so it applies to arbitrary geometry or lattice. 
The Hamiltonian $H$ (\ref{eq:ham}) does not need to be 1D or nearest-neighbor, that is,
the local gate $u$ can be of more general form.
Actually, arithmetic of Hamiltonian can cast a model $H$ into a 1D nearest-neighbor form,
as the latter form is universal for quantum computing~\cite{KPB+21}. 
This means that our EP method serves as a universal scheme for Hamiltonian quantum simulation.
Nevertheless, in order to maintain the geometric locality for high-dimensional systems,
extensions of MPS can be used, such as PEPS~\cite{SCP10}. 
For our purpose, we modify the PEPS form by requiring each local tensor 
being equivalent to a channel. 
For instance, on a 2D square lattice, a five-leg tensor $A_{ijkls}$ shall be a bipartite channel 
with input $i$ and $j$, output $k$ and $l$, and the physical index $s$.
A time flow can be consistently chosen among the channels forming the PEPS,
which is again a special application of our method.

% Here we develop a new method to study MPS dynamics, 
% which is in particular more suitable for quantum computers. 
% Our method relies on quantum channel (QC) dynamics instead of TC.
% The most primary principle emerging from this is the entanglement picture.

% this use space-time duality. 
% The space direction, which is the system size, maps to time. 
% The time direction, which is the circuit evolution, maps to space. 
% Namely, the physical system now is ``edge'' system that carries entanglement.
% Each local gate maps to a (sequence of) quantum channel evolution of an edge particle.

The scheme we developed can be applied to simulate more general systems,
such as models in quantum field theories. 
A model Hamiltonian can be mapped to a desirable form for quantum simulation by
mapping field operators to qubit operators,
or using equivalent first-quantized form and lattice discretization.
Also quantum field theory is nowadays more often considered as effective theory
for describing the low-energy physics of quantum many-body system~\cite{Wen04},
and it can capture the universal features of phases of matter
and phase transitions. 

A direct physical connection between MPS and quantum field theory is drawn through the 
continuous MPS theory~\cite{VC10,OEV10,DHO13}. 
For static features, it can be computed through a continuous-time dynamics described by a 
master equation of the edge system.
For dynamical features, the standard discrete MPS form is still proper,
so one merely needs to use discretization to map the model into a many-body form.

% EP method is more proper for systems with exponential growing dimension of Hilbert space.
% For other cases with a large Hilbert space dimension, 
% such as continuous-variable system, we can use discretization to map it to a qudit of dimension $d$,
% and encode a qudit in $n$ qubits for $d=2^n$.
% This applies to sparse Hamiltonian models such as Schr\"odinger equation $H=H_0+V(x)$, 
% and continuous-time quantum walk on a graph (lattice).

%\subsection{Quantum field theory}

% the key feature of QFT is their symmetry, different from non-relativistic models

\subsection{Thermal physics}

%Unitary channel case, there is no physical index. 

%Wick rotation

\begin{figure}
    \centering
    \includegraphics[width=0.4\textwidth]{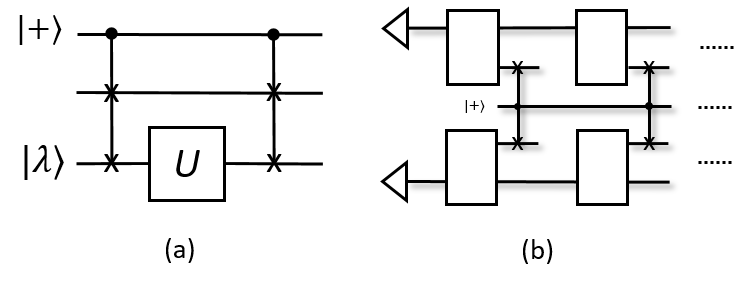}
    \caption{The DQC1 algorithm (a) and its combination with the EP scheme (b).
    The controlled-gate $\wedge_U$ is realized by the controlled-swap scheme
    with a qubit controller at $|+\ket$.
    The 2nd register carries the observable $A$,
    and the 3rd register is an eigenstate of $U$,
    which could be a ground state of $H$.
    Using EP, the controller with encoded $|+\ket_L$ executes the controlled-swap gates 
    on two MPSs, and 
    the implementation of $U$ is realized by a channel network.}
    \label{fig:dqcep}
\end{figure}

We now apply the EP method to study thermodynamics,
which also shows how EP can be employed as a module for solving complex problems. 
The primary task is to compute the thermal value 
\be \bra A \ket_\beta=\Tr (A e^{-\beta H}) \ee
of an observable $A$ for a model Hamiltonian $H$ and inverse temperature $\beta$. 
Our method is as follows. 
First, without loss of generality, 
we can choose $A$ to be diagonalizable with real eigenstates,
and the task reduces to the computation of $\bra a| e^{-\beta H} |a\ket$ for each $|a\ket$.
Using Taylor expansion $e^{-\beta H}=\sum_{n=0}^\infty \frac{(-\beta)^n}{n!} H^n$,
it further reduces to $\bra a| H^n |a\ket$ for all the orders $n$.
Now using Wick's rotation, 
we map temperature to time and consider $\bra a| U |a\ket$ for $U=e^{-it H}$.
This value can be computed by the DQC1 algorithm~\cite{KL98},
also known as Hadamard test,
whose circuit is shown in Fig.~\ref{fig:dqcep}.

On the input $P_+\otimes(P_a\otimes\pi)\otimes P_\lambda$
for $|\lambda\ket$ as an eigenstate of $U$, $\pi$ as a completely mixed state,
and $P_\psi:=|\psi\ket\bra\psi|$,  
we measure $\sigma_x \otimes P_a$ and $\sigma_y \otimes P_a$ 
to obtain two probabilities $p_x$ and $p_y$, respectively,
for Pauli operators $\sigma_x$ and $\sigma_y$ of the controller.
From two 2nd order equations satisfied by $p_x$ and $p_y$,
it is easy to solve the value $\bra a| U |a\ket$ for each eigenstate $|a\ket$. 

To compute $\bra a| H^n |a\ket$ with a finite truncation order
in the 
Taylor expansion $e^{-it H}=\sum_{n=0}^{s-1} \frac{(-it)^n}{n!} H^n +O(t^s)$,
we need to simulate $U$ for various time parameters. 
For a model $H=\sum_r H_r$, we use the 1st order Trotter sequence 
\be T(t)=\prod_r e^{-it H_r}= e^{-it H}+ O(t^2), \ee 
and for $t=R\tau$, $(T(\tau))^R= e^{-it H}+ O(R\tau^2)$.
Given a decomposition accuracy parameter $\epsilon$ and a Taylor truncation order $s$,
we can find the parameters $R$ and $\tau$ to realize $(T(\tau))^R$.
With accuracy $\C O(\epsilon)$ for each $\bra a| H^n |a\ket$,
this finally computes $\bra A \ket_\beta$ with the same order of accuracy. 
A merit of the Taylor expansion is that 
the truncation converges fast as the order 
$s\in \C O(\log \frac{1}{\epsilon})$~\cite{BCC15}.
The Trotter decomposition can be extended to higher-order forms to improve the simulation~\cite{BACS07}.
The simulation cost mainly includes the controlled-swap gates 
together with the cost for each $U$ up to the truncation order $s$.

To realize the algorithm on quantum processors, 
there is a notable difference for photonic platforms and solid-state platforms.
The controlled-gate $\wedge_U$ can be realized via a 
Mach-Zehnder interferometer without using controlled-swap gates
since there is a direct-sum structure of the Hilbert space of photonic qubits~\cite{AFC14}.
When controlled-swap gates are implemented for solid-state platforms,
the circuit depth for the qubit controller is large, 
and encoding can be used to reduce the potential effects of errors.
For instance, the repetition code will convert $|+\ket$ into a GHZ state with logical state
$|+\ket_L=\frac{1}{\sqrt{2}}(|00\cdots 0\ket+|11\cdots 1\ket)$,
and each qubit only executes two controlled-swap gates 
before and after the implementation of the evolution $U$. 
More advanced codes can also be employed with an error-correction overhead.
Generically, it would depend on the details of an implementation 
to choose a proper scheme for the controller,
as well as the realizations of controlled-swap gates.

\subsection{Extensions}

We consider a few extensions of the above algorithms. 
If the Hamiltonian being considered is a so-called entanglement Hamiltonian~\cite{ZCZ+15},
then our algorithms can be used to compute entropy. 
Namely, for an entropy $S(\rho)=-\Tr \rho \log \rho$ of a state $\rho$,
which can be a local part of a whole system or a state on its own,
a modular entanglement Hamiltonian $H$ is defined such that $\rho= e^{-H}$,
wherein a temperature parameter is absorbed in $H$ itself. 
Then the entropy value is expressed as 
\be  S(\rho)=\Tr (H e^{-H} ). \ee
For local $H=\sum_r H_r$, the above can be computed for each term 
$\Tr (H_r e^{-H} )$ with our algorithm, and sum up to obtain $S(\rho)$.
The entropy $S(\rho)$ is a nonlinear function of the state $\rho$,
which is usually not easy to obtain or measure directly.
Notably, our algorithm can compute $S(\rho)$ 
rather than its R\'enyi entropy extensions~\cite{B19},
which do not require all moments $\rho^n$ of the state $\rho$. 
Being able to measure $S(\rho)$, 
it is also straightforward to study its time evolution.

Computationally, the schemes above compute values $|\bra \phi|\psi\ket|^2$
and $\bra \psi|U|\psi\ket$ for given pure states $|\psi\ket$, $|\phi\ket$, 
and unitary gate $U$.
We can consider more general quantity in the form of $\bra \phi|U|\psi\ket$,
which matters in broad physical contexts, e.g., 
in the out-of-time-ordered-correlation studied in non-equilibrium physics~\cite{OTOC23}
and the coefficients in the operator-product expansion in conformal field theory~\cite{FMS97},
and also it is the analog of the propagator $\bra x_b | e^{-iH (t_b-t_a)}|x_a\ket$ 
that plays a central role in path integral. 

Given a known state $|\psi\ket$, we define the reflection operator 
$R_\psi:=\I - 2 |\psi\ket \bra \psi|$, and the value 
$\bra \phi|U|\psi\ket$ is deduced from the value 
\bea &\bra 0| R_\phi U R_\psi |0\ket = & \bra 0| U |0\ket 
+ 4 \bra 0| \phi\ket \bra \phi|U|\psi\ket \bra \psi |0\ket
 \\ \nonumber
&& - 2 \bra 0| \phi\ket \bra \phi |U |0\ket  - 2 \bra 0|U |\psi\ket \bra \psi |0\ket , 
\eea 
which can be computed by the DQC1 algorithm,
wherein the values $\bra 0| U |0\ket $, $\bra 0| \phi\ket$, and $\bra \psi |0\ket$
are easy to obtain for $|0\ket$ denoting a computational basis state.
The value $\bra \phi |U |0\ket$ and $\bra 0|U |\psi\ket$ is deduced from  
$\bra 0| R_\phi U |0\ket$ and $\bra 0| U R_\psi |0\ket$, respectively,
which are also DQC1 computable.  
Note here for a product, say, $R_\phi U$, 
there is no need to use an eigenstate of it;
instead, an eigenstate of $U$ and an eigenstate of $R_\phi$, 
which can be the state $|\phi\ket$ itself, 
suffices to construct the DQC1 algorithm.
The last ingredient in our method is the implementation of a reflection operator,
which also plays a central role in Grover's search algorithm~\cite{Gro96}.
As a state $|\psi\ket$ is known, we can find a gate $U_\psi$ so that 
$|\psi\ket=U_\psi |0\ket$, then 
\be R_\psi=U_\psi (\I-2|0\ket\bra0|) U_\psi^\dagger, \ee
which means $R_\psi$ is constructed from $U_\psi $ and a multiple-controlled phase 
gate $\I-2|0\ket\bra0|$.

For more general values $\bra \phi|A|\psi\ket$ with a matrix $A$ 
that is not unitary, they can be computed by first decomposing $A$ as a 
combination of a few unitary matrices $A=\sum_i a_i U_i$,
and then compute each term $\bra \phi|U_i|\psi\ket$ on a quantum computer. 
The first step is assumed to be classical and there are many methods to achieve this.
For instance, by considering the traceless version $A-\I \text{tr}A$ instead of $A$ itself
and renormalizing $\|A\|\leq 2$,
it can be expressed as a sum of two unitary matrices $A=U_++U_-$ for $U_\pm= B\pm iC$,
with $B=A/2$, 
$C=\sqrt{\I-B^\dagger B}$.
The matrix $C$ can be rather easily obtained as the square root of a nonnegative 
semidefinite matrix.

% \subsection{Entanglement entropy}

% how to measure EE? there are method to measure Renyi 2nd order purity, 
% $tr(\rho^2)$, which can be measured by SWAP test, which is a special case of DQC1,
% or by classical shadow random PVM scheme. 

% is there any macroscopic quantity that depends on EE? 

% for CFT, the central charge relates to EE. 

% we aim to consider time-evolution $S(\rho)(t)$.

% \section{Open task: what else can we do?}

% \begin{itemize}
%     \item Semi-classical limit: is there a way to separate what `part'
%     is classically efficiently simulatable?  (via tensor-network state);
%     \item what other problems we can consider?
%     \item I know gapless state, relate to MERA, any problem to study? 
% \end{itemize}

% \section{Extensions}

% 1) if the entanglement system is available on the first hand, 
% we can do more general operations, and these are quantum superchannels

\section{Discussion and Conclusion}

In this work, we proposed a new picture in quantum mechanics, 
named the entanglement picture (EP), which is inspired by the fundamental concept of 
quantum entanglement. We showed that it is natural to describe quantum dynamics
within this framework, especially for the purpose of quantum computing,
which often involves the processing of a large amount of entanglement. 
The quantum algorithms we developed serve as illustrations of the usage of EP,
though it shall be noted that they may not be optimal for computing any particular quantity.

The EP is based on a fundamental principle in quantum mechanics: 
the channel-state duality. This duality is as important as entanglement, and they are intimately related:
quantum dynamics on a system must be completely positive (i.e., described by a channel)
rather than merely positive, precisely because of entanglement with its environment.
We have shown that this duality underlies the bulk-edge duality 
of many-body entangled states, namely, matrix-product states (MPS). 

Using EP, we describe quantum dynamics as a network of channels,
which can be used to construct quantum simulation algorithms for computing overlaps.
The algorithms are suitable for generic local Hamiltonians and observables,
and hence can be used for a wide range of problems. 
The quantum circuits for the algorithms are deterministic,
and the sampling costs arise from classical processing 
and the estimation of outcome probabilities. 
It remains to be seen whether there could be a certain computational
or engineering advantage for using the EP method 
compared with standard methods developed for specific problems.

Nevertheless, we can identify several scenarios where EP may be particularly suitable. 
First, for high-entanglement dynamics beyond classical reach, 
when the bond dimension grows exponentially (e.g., during thermalization or long-time evolution), 
classical methods like DMRG become intractable, while EP executed on a quantum device 
can in principle access such states. 
Second, for observable-specific simulation, 
unlike classical tensor network methods that typically require a global contraction 
of the state representation to compute local observables, 
EP can target specific expectation values more directly, 
enabling potentially more efficient use of quantum resources. 
Third, for distributed quantum computing, EP naturally decomposes a large quantum circuit 
into smaller fragments that can be executed on separate quantum processors, 
as the binary Bell measurement we use can be viewed as a circuit-knitting or fusion scheme 
within the broader paradigm of modular computing~\cite{CAF24,BCD25,XLS25,PS23,UPR23,LMH23}.
That said, the quantum circuit model remains the standard setting for designing quantum algorithms, 
and for an arbitrary circuit—especially those with low entanglement—there may be no benefit 
to using the language of MPS and EP.

Beyond quantum simulation, 
the EP perspective may offer insights into other areas of quantum information and computation~\cite{WM21}. 
For instance, EP is conceptually related to measurement-based quantum computing (MBQC), 
where entanglement is consumed by on-site measurements, 
while in EP the entanglement space serves as a resource that is processed by channel networks. 
Another universal model that directly relates to MPS is the local quantum Turing machine~\cite{WT20}, 
where the entanglement system acts as the `memory' and operations are applied directly on it 
rather than on the physical space. 
Such operations can be generalized using superchannels to generate higher-order MPS 
with larger bond dimensions~\cite{CDP08}. 
These connections suggest that EP may provide a unified language for understanding 
different models of quantum computation through the lens of entanglement dynamics.

To conclude, roughly speaking, the entanglement picture is a network of quantum channels. 
The framework of EP is interesting for several reasons:
it adds a new picture to work with in quantum mechanics, 
it further illustrates the fundamental importance of entanglement,
and it broadens the range of tools from quantum information 
that can be applied to solve problems in other fields of physics.

\section{Acknowledgement}
This work has been funded by
the National Natural Science Foundation of China under Grants
12447101 and 12105343.

% \newpage 
%\bibliographystyle{alpha}
%\bibliography{ext}

\bibliography{ext}{}

\begin{thebibliography}{10}
\expandafter\ifx\csname url\endcsname\relax
  \def\url#1{\texttt{#1}}\fi
\expandafter\ifx\csname urlprefix\endcsname\relax\def\urlprefix{URL }\fi
\expandafter\ifx\csname href\endcsname\relax
  \def\href#1#2{#2} \def\path#1{#1}\fi

\bibitem{HHH09}
R.~Horodecki, P.~Horodecki, M.~Horodecki, K.~Horodecki, Quantum entanglement,
  Rev. Mod. Phys. 81 (2009) 865--942.

\bibitem{NC00}
M.~A. Nielsen, I.~L. Chuang, Quantum Computation and Quantum Information,
  Cambridge University Press, Cambridge U.K., 2000.

\bibitem{RT06}
S.~Ryu, T.~Takayanagi, Holographic derivation of entanglement entropy from the
  anti--de sitter space/conformal field theory correspondence, Phys. Rev. Lett.
  96 (2006) 181602.

\bibitem{ZCZ+15}
B.~Zeng, X.~Chen, D.-L. Zhou, X.-G. Wen, Quantum Information Meets Quantum
  Matter, Springer-Verlag New York, 2019.

\bibitem{PVW+07}
D.~Perez-Garcia, F.~Verstraete, M.~Wolf, J.~Cirac, Matrix product state
  representations, Quantum Information \& Computation 7~(5) (2007) 401--430.

\bibitem{Cho75}
M.-D. Choi, Completely positive linear maps on complex matrices, Linear Algebra
  Appl. 10 (1975) 285--290.

\bibitem{Jam72}
A.~Jamio{\l}kowski, Linear transformations which preserve trace and positive
  semidefiniteness of operators, Rep. Math. Phys. 3 (1972) 275.

\bibitem{Fey48}
R.~P. Feynman, Space-time approach to non-relativistic quantum mechanics, Rev.
  Mod. Phys. 20 (1948) 367--387.

\bibitem{PKS19}
S.~Paeckel, T.~Köhler, A.~Swoboda, S.~R. Manmana, U.~Schollwöck, C.~Hubig,
  Time-evolution methods for matrix-product states, Annals of Physics 411
  (2019) 167998.

\bibitem{AKLT87}
I.~Affleck, T.~Kennedy, E.~H. Lieb, H.~Tasaki, Rigorous results on valence-bond
  ground states in antiferromagnets, Phys. Rev. Lett. 59 (1987) 799--802.

\bibitem{RB01}
R.~Raussendorf, H.~J. Briegel, A one-way quantum computer, Phys. Rev. Lett. 86
  (2001) 5188--5191.

\bibitem{BACS07}
D.~W. Berry, G.~Ahokas, R.~Cleve, B.~C. Sanders, Efficient quantum algorithms
  for simulating sparse hamiltonians, Commun. Math. Phys. 270~(2) (2007)
  359--371.

\bibitem{JLP12}
S.~P. Jordan, K.~S.~M. Lee, J.~Preskill, {Quantum algorithms for quantum field
  theories.}, Science 336~(6085) (2012) 1130--3.

\bibitem{BBM20}
B.~Bauer, S.~Bravyi, M.~Motta, G.~K.-L. Chan, Quantum algorithms for quantum
  chemistry and quantum materials science, Chemical Reviews 120~(22) (2020)
  12685--12717.

\bibitem{WC20}
D.-S. Wang, Choi states, symmetry-based quantum gate teleportation, and
  stored-program quantum computing, Phys. Rev. A 101 (2020) 052311.

\bibitem{WQ22}
D.-S. Wang, {A prototype of quantum von Neumann architecture}, Commun. Theor.
  Phys. 74 (2022) 095103.

\bibitem{BZ06}
I.~Bengtsson, K.~\.{Z}yczkowski, Geometry of Quantum States, Cambridge
  University Press, Cambridge U.K., 2006.

\bibitem{Vid04}
G.~Vidal, Efficient simulation of one-dimensional quantum many-body systems,
  Phys. Rev. Lett. 93 (2004) 040502.

\bibitem{XLS25}
X.~Xu, Y.-D. Liu, S.~Sha, Y.-J. Wang, D.-S. Wang, Distributed quantum computing
  with black-box subroutines, Quantum Sci. Technol. 10 (2025) 045014.

\bibitem{W15}
D.-S. Wang, Weak, strong, and uniform quantum simulations, Phys. Rev. A 91
  (2015) 012334.

\bibitem{BCC+14}
D.~W. Berry, A.~M. Childs, R.~Cleve, R.~Kothari, R.~D. Somma, Exponential
  improvement in precision for simulating sparse hamiltonians, in: Proc. 46th
  ACM Symposium on Theory of Computing, 2014, p. 283.

\bibitem{BHM02}
G.~Brassard, P.~Hoyer, M.~Mosca, A.~Tapp, Quantum amplitude amplification and
  estimation, Contem. Mathemat. 305 (2002) 53–74.

\bibitem{Hay17}
M.~Hayashi, Quantum Information Theory: Mathematical Foundation, 2nd edition,
  Springer, 2017.

\bibitem{GKS05}
D.~Gobert, C.~Kollath, U.~Schollw\"ock, G.~Sch\"utz, Real-time dynamics in
  spin-$\frac{1}{2}$ chains with adaptive time-dependent density matrix
  renormalization group, Phys. Rev. E 71 (2005) 036102.

\bibitem{KPB+21}
T.~Kohler, S.~Piddock, J.~Bausch, T.~Cubitt, General conditions for
  universality of quantum hamiltonians, PRX Quantum 3 (2022) 010308.

\bibitem{SCP10}
N.~Schuch, J.~I. Cirac, D.~Perez-Garcia, {PEPS as ground states: Degeneracy and
  topoloy}, Ann. Phys. 325 (2010) 2153.

\bibitem{Wen04}
X.-G. Wen, Quantum field theory of many-body systems, Oxford University Press,
  2004.

\bibitem{VC10}
F.~Verstraete, J.~I. Cirac, Continuous matrix product states for quantum
  fields, Phys. Rev. Lett. 104 (2010) 190405.

\bibitem{OEV10}
T.~J. Osborne, J.~Eisert, F.~Verstraete, Holographic quantum states, Phys. Rev.
  Lett. 105 (2010) 260401.

\bibitem{DHO13}
D.~Draxler, J.~Haegeman, T.~J. Osborne, V.~Stojevic, L.~Vanderstraeten,
  F.~Verstraete, Particles, holes, and solitons: A matrix product state
  approach, Phys. Rev. Lett. 111 (2013) 020402.

\bibitem{KL98}
E.~Knill, R.~Laflamme, Power of one bit of quantum information, Phys. Rev.
  Lett. 81 (1998) 5672--5675.

\bibitem{BCC15}
D.~W. Berry, A.~M. Childs, R.~Cleve, R.~Kothari, R.~D. Somma, Simulating
  hamiltonian dynamics with a truncated taylor series, Phys. Rev. Lett. 114
  (2015) 090502.

\bibitem{AFC14}
M.~Araujo, A.~Feix, F.~Costa, C.~Brukner, Quantum circuits cannot control
  unknown operations, New J. Phys. 16 (2014) 093026.

\bibitem{B19}
T.~Brydges, et~al., {Probing Rényi entanglement entropy via randomized
  measurements}, Science 364 (2019) 260.

\bibitem{OTOC23}
I.~García-Mata, R.~A. Jalabert, D.~A. Wisniacki, Out-of-time-order correlators
  and quantum chaos, Scholarpedia 18 (2023) 55237.

\bibitem{FMS97}
P.~Francesco, P.~Mathieu, D.~Sénéchal, Conformal field theory,
  Springer-Verlag New York, 1997.

\bibitem{Gro96}
L.~K. Grover, A fast quantum mechanical algorithm for database search, in:
  Proceedings of the twenty-eighth annual ACM Symposium on Theory of Computing,
  1996.

\bibitem{CAF24}
M.~Caleffi, M.~Amoretti, D.~Ferrari, et~al., Distributed quantum computing: a
  survey, Computer Networks 254 (2024) 110672.

\bibitem{BCD25}
D.~Barral, F.~J. Cardama, G.~Díaz-Camacho, et~al., Review of distributed
  quantum computing: From single {QPU} to high performance quantum computing,
  Computer Science Review 57 (2025) 100747.

\bibitem{PS23}
C.~Piveteau, D.~Sutter, Circuit knitting with classical communication, IEEE
  Trans. Inform. Theory 70 (2023) 3310797.

\bibitem{UPR23}
C.~Ufrecht, M.~Periyasamy, S.~Rietsch, D.~D. Scherer, A.~Plinge, C.~Mutschler,
  Cutting multi-control quantum gates with zx calculus, Quantum 7 (2023) 1147.

\bibitem{LMH23}
A.~Lowe, M.~Medvidović, A.~Hayes, et~al., Fast quantum circuit cutting with
  randomized measurements, Quantum 7 (2023) 934.

\bibitem{WM21}
D.-S. Wang, A comparative study of universal quantum computing models: towards
  a physical unification, Quantum Engineering 2 (2021) 85.

\bibitem{WT20}
D.-S. Wang, A local model of quantum {T}uring machines, Quant. Infor. Comput.
  20~(3) (2020) 0213--0229.

\bibitem{CDP08}
G.~Chiribella, G.~M. D'Ariano, P.~Perinotti, Quantum circuit architecture,
  Phys. Rev. Lett. 101 (2008) 060401.

\end{thebibliography}
\bibliographystyle{elsarticle-num}

\end{document}